\pdfoutput=1

\documentclass[aps,preprintnumbers,amsmath,amssymb,twocolumn, tightenlines,superscriptaddress]{revtex4}

\usepackage{graphicx}
\usepackage{epsfig}

\newcommand{\be}{\begin{eqnarray}}
\newcommand{\ee}{\end{eqnarray}}

 \newcommand{\gsim}{\mathrel{\hbox{\rlap{\lower.55ex \hbox {$\sim$}}
                   \kern-.3em \raise.4ex \hbox{$>$}}}}
\newcommand{\lsim}{\mathrel{\hbox{\rlap{\lower.55ex \hbox {$\sim$}}
                   \kern-.3em \raise.4ex \hbox{$<$}}}}

\newcommand{\ba}{\begin{eqnarray}}
\newcommand{\ea}{\end{eqnarray}}

\begin{document}


\title{Effect of Light Fermions on the Confinement Transition in QCD-like Theories }
\author {Jinfeng Liao} 
\address{Physics Department and Center for Exploration of Energy and Matter,
Indiana University, 2401 N Milo B. Sampson Lane, Bloomington, IN 47408, USA.}
\address{RIKEN BNL Research Center, Bldg. 510A, Brookhaven National Laboratory, Upton, NY 11973, USA.}
\author{Edward Shuryak} 
\address{Department of Physics and Astronomy, State University of New York,
Stony Brook, NY 11794, USA.}

\begin{abstract}
Dependence of the confinement transition parameters on the fermion content provides information on the mechanism of confinement. Recent progress in lattice gauge theories has allowed to study it 
for light flavor number $N_f\sim O(10)$ and found this transition to shift toward significantly stronger coupling. We propose an explanation for that:  light fermions can occupy  the chromo-magnetic monopoles, via zero modes,  making them ``distinguishable" and unsuitable for Bose-Einstein Condensation. Such dilution of unoccupied monopoles  is compensated by stronger coupling that makes them lighter  and more numerous. We also suggest that flavor-carrying quark-monopole objects account 
for the density beyond quark Fermi sphere seen in cold dense phase of  $N_c=2$ lattice QCD.
\end{abstract}

\maketitle

1. The color confinement of Quantum Chromodynamics (QCD) remains to be one of the most outstanding puzzles of the  Standard Model, in spite of intense
studies  via lattice QCD simulations \cite{confinement_review}.  By virtue of the asymptotic freedom, 
at high temperature the effective coupling is weak and the matter is in a deconfined
phase known as the quark-gluon plasma, or QGP \cite{Shuryak:1978ij}. When $T$ is lowered, as has happened after the Big Bang
or in heavy ion collisions, the effective coupling grows. At certain critical temperature $T_c$, 
a transition into the confined hadronic world occurs\cite{Shuryak_review,Gyulassy_McLerran}. 
Physics had many examples in which the dependence of $T_c$ on some parameters
had offered crucial insights: e.g.  when the $T_c$ of superconductivity
had shown an  isotope mass dependence. 
 Our strategy is similar: examining how this transition depends on the fermion representation and flavor number $N_f$ may lead to insights about the mechanism of confinement.

\begin{figure}[!t]\vskip 0.025in
 \center{ \hskip 0in\includegraphics[width=6.6cm]{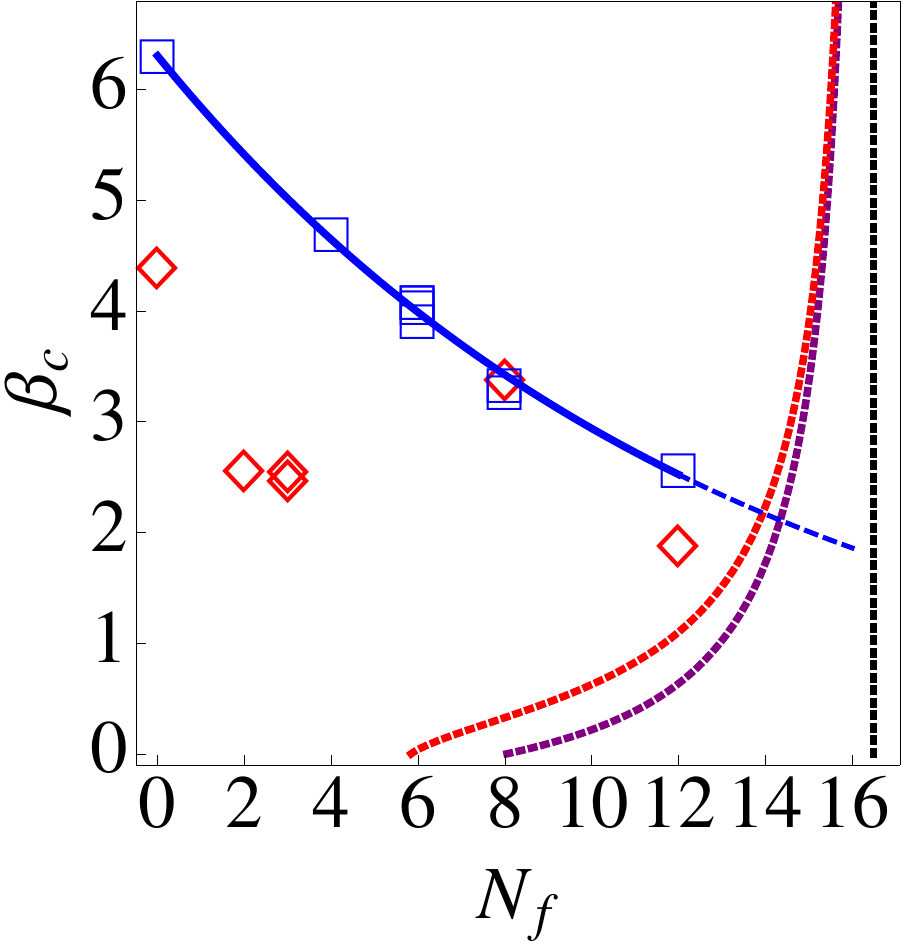}}
 \caption{\label{fig_fundamentals} Dependence of the critical lattice coupling $\beta_c$ at scale $T_c$ versus the number of fundamental quark flavors $N_f$ in QCD-like theories.  Blue boxes are from \cite{Miura:2011mc}  with near-coincident boxes being lattice data for the same $N_f$ with different $N_\tau$ which demonstrate lattice spacing consistency. Red diamonds are from various other literature. The thick blue line is the fitting curve, extended as dashed blue line beyond $N_f=12$. The black/purple/red dashed curves on the right are lines for vanishing beta function at 1,2,3-loop levels.}
 \end{figure}

 Our phenomenological input comes from the lattice studies.  A (very incomplete) list of those ranges from 
 the well studied region of 
 $N_f=2,3$ (see e.g. \cite{Karsch:2000ps,Cheng:2006aj} to the recent extension toward $N_f=8$ \cite{Jin:2010vm,Deuzeman:2008sc} and even $N_f=12 $ \cite{Miura:2011mc}, which has attracted special attention
 in connection with the search for a conformal regime, see e.g. \cite{Fodor:2011tu}. There are also studies with adjoint \cite{Karsch:1998qj,Cossu:2009sq,DeGrand:2011qd} and tensor-symmetric  quarks 
 \cite{DeGrand:2009hu} in similar regime, see e.g. reviews \cite{DeGrand:2010ba,Lucini:2003zr}.  Starting from the
   pure gauge theory $(N_f=0)$ and increasing $N_f$,  one finds the monotonous and persistent shift toward the 
   stronger coupling at $T_c$. (The value of $T_c$ itself is usually expressed via units customary
   in lattice community, which fixes the $T=0$  string tension to the same real-world value. 
We will not use 
such units as they confuse the comparison across different QCD-like theories.)
The absolute magnitude of the gauge coupling constant $\beta_c$ at the transition temperature $T_c$, instead, bears more direct information. For doing so, we evolve   the critical lattice coupling $\beta_L=2Nc/g^2$ at the lattice  scale $a$  (by 2-loop running) to the scale $1/T_c=N_\tau a \, $:
\begin{eqnarray}
\left(\frac{2N_c}{\beta_c}\right)^{-\frac{b_1}{b_0^2}}\, e^{-\frac{(4\pi)^2}{2N_c b_0} \beta_c}
 = N_\tau^2 \, \left(\frac{2N_c}{\beta_L}\right)^{-\frac{b_1}{b_0^2}}\, e^{-\frac{(4\pi)^2}{2N_c b_0} \beta_L} \,\,
\end{eqnarray}
where $b_0,b_1$ are the usual $\beta$-function coefficients. In Fig.\ref{fig_fundamentals} we've collected such $\beta_c$ values for theories with varied $N_f$ from various 
lattice simulations. Although the qualitative trend is clear, quantitatively it is still hard to compare the works of different lattice groups even for the same theory due to e.g. difference in the actions used.  Only the recent data from $N_f=0$ to $N_f=12$ from the same group \cite{Miura:2011mc} (shown as blue boxes in Fig.\ref{fig_fundamentals})  allow for direct  comparison. (These results are for chiral restoration, which for the fundamental quarks is believed to trace the deconfinement rather closely.) The main observation from Fig.\ref{fig_fundamentals} is  that the critical coupling $\beta_c$ changes by a substantial factor  when the flavor number increases from $N_f=0$ to $N_f=12$. Similarly,
the $N_c=2$  theory with two adjoint fermions flows into the coupling at the infrared fixed point  $1/g^2_*=0.20 (4) (3)$ \cite{DeGrand:2011qd}, about a
factor 4 stronger than the critical one at the deconfinement of the $N_c=2$ pure gauge theory.

 2. Let us now turn to the mechanism of confinement.  't Hooft and Mandelstam suggested  a ``dual superconductor'' model  \cite{'t Hooft-Mandelstam}  related it to the
   Bose-Einstein Condensation (BEC) of  chromo-magnetic monopoles.
  For reviews of those ideas at $T=0$ in lattice
  context see e.g. \cite{confinement_review}. Only recently it was realized that 
  if monopoles are indeed the emergent excitations 
  one should better study them in a ``normal" phase. Furthermore, in
  the region  right above the transition $T>T_c$  such monopoles should be the dominant thermal degrees of freedom ``ready'' for BEC \cite{Liao:2006ry,Chernodub}. Lattice and model studies have shown that the effective coupling of the ``magnetic plasma" does run as $1/g$, inversely to the electric \cite{D'Alessandro:2007su,Unsal}. 
 Consequences of the magnetic scenario help  to understand other lattice results   \cite{Liao:2005hj}  as well as heavy ion experiments  \cite{Liao:2008dk,Ratti:2008jz}. This study is based on  the scenario that confinement occurs as Bose condensation of monopoles that are the dominant physical degrees of freedom in the plasma near $T_c$. 

 How can the light fermions affect the monopoles? It is known that the light fermions can become attached to them.  The so called ``fermionic zero modes"  
 are specific bound states in which positive kinetic energy of localization exactly cancels
  the magnetic-moment-field interaction. (Their existence and number  $N_M$ are  required
 by the topological index theorems, and thus insensitive to any perturbative monopole deformations.)  While such states may still be bosons, they carry flavor indices due to the fermions and are ``distinguishable", thus not contributing to the BEC of ``unoccupied" monopoles.

These zero modes are known explicitly for 't Hooft-Polyakov monopoles \cite{'t Hooft-Polyakov} which are present in gauge theories with adjoint scalars, such as in ${\cal N}=2$ supersymmetric gauge theories. Since flipping the charge and the spin leads to the same Dirac equation,
antiquarks also have the same zero modes. Furthermore,
 for each of these zero mode states, it  can be either populated or not, so the number of totally available states grows exponentially $\sim 2^{2 N_f N_M}$. 
Spectroscopy of those states in the supersymmetric  setting was developed in 1990's, see e.g.
 \cite{Harvey:1996ur}. ``Magnetic supermultiplets"  have been explicitly checked for two
famous conformal theories,  the $\cal N$=4 SYM and  the $\cal N$=2 SQCD with $N_f=4$: in bot cases 
one finds exactly the same set of spins/multiplicities
as that in the original  electric one. (Since those theories are electric-magnetic self-dual,
their coupling cannot run at all!)  
There is one zero mode for the fundamental while two for the adjoint (Dirac) fermions:  so these monopole-single-fermion states case into spin $0$ and $\frac{1}{2}$ objects, respectively.
 While in the static case zero mode states  are degenerate with the pure monopole, it is not so for the dynamical lattice monopoles with non-static paths.

3. The BEC criteria for an interacting boson ensemble was proposed by  Feynman 50 years ago for the study of liquid $^4 He$ \cite{Feynman}
 and was recently generalized by the analysis of Cristoforetti and Shuryak \cite{Cristoforetti:2009tx}. In the  finite-$T$ description  with periodic paths there appear ``$k$-clusters" of bosons interchanging their initial (at Matsubara time $0$) and final (at Matsubara time $\beta=1/T$) positions.
 Those clusters can be depicted as ``Feynman polygons" with $k$ points.  Their probability depends on two competing factors, suffering from  suppression  due to the extra action  $exp(-kS_{ex})$   while benefiting from large combinatory number of $k$-polygons.  The balance point marks the onset of condensation by divergence of the sum over the $k$-clusters. This method has been used
 in \cite{D'Alessandro:2010xg} where it was shown that  lattice monopoles do have the BEC transition exactly at $T_c$. See also other studies of macroscopic clusters  (or ``percolation'') such as\cite{Chernodub,D'Alessandro:2007su}.
  According to Feynman, quantitative BEC condition is a $universal$ critical value of the extra action per particle, which in three spatial dimensions is given by 
\begin{eqnarray} \label{eq_sc}
S_{ex} \le S_c\approx 1.655
\end{eqnarray}
Upon fulfilling (\ref{eq_sc}), long sequences of ``hopping'' bosons will occur, creating a macroscopic ``supercurrent". Its validity for interacting systems is demonstrated in \cite{Cristoforetti:2009tx}.

Let us now apply the above criteria to the monopole condensation in QCD-like theories. The minimal exchange action $S_{ex}$ for two nearest-neighbor bosons that are separated by a typical distance $d=n^{-1/3}$ (with $n$ the number density) during the Matsubara time from $\tau=0$ to $\tau=\beta=1/T$ could be estimated as
 \begin{eqnarray}
S_{ex}= m^* \sqrt{\beta^2+d^2} - m^* \beta +S_V 
\end{eqnarray}
with an explicitly written kinetic term,
containing an effective mass $m^*$, and the implicit potential term $S_V$. When close to condensation, the monopoles 
are very dense, with typical spatial separation $d$ comparable or smaller than the inverse temperature $\beta$, therefore justifying 
a further approximation of the above expression: 
$S_{ex} \approx \frac{1}{2} m^* T d^2 + S_V  =  \frac{1}{2} \left(\frac{m^*}{T}\right) \left(\frac{n}{T^3}\right)^{-{2\over 3}} + S_V$. 
The term $S_V=\int _0^{1/T} V(r(\tau))d\tau$ is basically the ratio of the inter-monopole interaction potential and $T$, also known as classical plasma coupling   $\Gamma_M \sim <V>/T \sim S_V$. As shown in Fig.3b of \cite{Liao:2006ry},  at high $T$ where the
``magnetic scaling" $d\sim 1/(g^2 T)$ and $g_{magnetic}\sim 1/g$ works, this ratio $\Gamma_M$ does not depend on the coupling or T and is a constant $\approx 5$, while close to condensation it decreases to a value of about 1 as $T\to T_c$ for $N_f=0$.  The following onset condition will then be applied for the monopole condensation in QCD-like theories for the rest of our analysis: 
\begin{eqnarray} \label{Eq_condition}
 \left(\frac{m^*}{T}\right) \left(\frac{n}{T^3}\right)^{-{2\over 3}} \le \tilde{S}_c \equiv 2\, (S_c-S_V)
\end{eqnarray}
The constant $\tilde{S}_c$ is of order one and its precise value is not needed as long as its $N_f$-dependence is negligible. 

4. Now, how would the transition get affected by adding light fermions? As already pointed out, the fermions can be attached to some of the monopoles via zero modes and effectively reduce the number of identical monopoles. Consider a monopole with one flavor of light quark added: for each of its allowed zero modes there is a probability for it to be occupied by a fermion or not. Let us assume the ratio of the probabilities (occupied/not-occupied) to be $f$ (a kind of zero-mode fugacity), we then see that the overall probability for a monopole (with $N_M$ number of zero modes for {\em each fermion flavor}) to stay as a ``pure'' monopole is simply $1/(1+f)^{2 N_M}$ (with the factor 2 accounting for both quark and anti-quark contributions for Dirac fermions). So effectively the available density for BEC condensation will be $n/(1+f)^{2 N_f N_M}$. Combined with the BEC condition in Eq.(\ref{Eq_condition}) we obtain
\begin{eqnarray} \label{Eq_flavor}
\left(\frac{m^* }{ T}\right)\, \frac{(1+f)^{4 N_f N_M/3}}  { (n/T^3)^{2/3} }   \le \tilde{S}_c
\end{eqnarray} 
This implies that with increasing $N_f$, the density $n$ has to increase and mass $m^*$ to decrease correspondingly so as to reach the same condensation condition. This pushes the transition to stronger coupling, therefore explaining the $N_f$-dependence of the critical coupling in Fig.\ref{fig_fundamentals}. 

To make a semi-quantitative estimate, we use the following magnetic scaling relations that connect the monopole mass and density with gauge coupling: $m^*/T \sim 1/g$ and $n^{1/3}/T \sim g^2$ \cite{Polyakov,Liao:2006ry,D'Alessandro:2007su}. Combined with the above condition, we obtain the critical gauge coupling for monopole condensation to be:
$g (N_f) = g_0 \, (1+f)^{4 N_f N_M/15}$
 where $g_0$ is the corresponding critical coupling for pure gauge case. This can be further converted to the lattice couping $\beta=2N_c/g^2$:
\begin{eqnarray} \label{Eq_beta}
\beta_c (N_f) = \beta_0 (1+f)^{-8 N_f N_M/15}
\end{eqnarray}

As a concrete example let us focus on the case of fundamental fermions (with $N_M=1$). The observable we examine is the critical (lattice) gauge coupling, $\beta_c\equiv \beta(T=T_c)=2N_c/g^2(T_c)$ as a function of $N_f$, as shown in fig.\ref{fig_fundamentals}. In particular we concentrate on the data for $N_f=0$ to $N_f=12$ from the same group \cite{Miura:2011mc}, shown in Fig.\ref{fig_fundamentals} as blue boxes. We've used the above Eq.(\ref{Eq_beta}) to make a fit for these data (blue boxes) and obtained the optimal value $f\approx 0.154\,$: the fitting curve is shown as the thick blue line in Fig.\ref{fig_fundamentals}. Our model formula in Eq.(\ref{Eq_beta}) with one parameter nicely describes all the data points  in \cite{Miura:2011mc} from $N_f=0$ to $N_f=12$!  The suppression factor $f$ of monopole-quark as compared with pure monopole may be understood as follows: the monopole-quark has color-electric charge, and in the near-$T_c$ plasma it was known from previous studies \cite{Liao:2005hj} that the electric particles are heavier than the magnetic particles by roughly $\Delta M \sim 2 T_c$, thus leading to a suppression factor $\sim e^{-\Delta M / T}\sim 0.135$ that is fairly close to  $f\approx 0.154$.

For completeness we've also displayed in Fig.\ref{fig_fundamentals} the lattice results from various other groups  \cite{Karsch:2000ps,Cheng:2006aj,Jin:2010vm,Deuzeman:2008sc,Fodor:2011tu}: these are shown as red diamonds. Admittedly, 
there are uncertainties due to different lattice actions and ambiguities associated with the possible differences between chiral/confinement transitions, which shall all be sorted out in the future lattice simulations. Nevertheless the qualitative trend of decreasing lattice coupling with $N_f$ is well in line with the data in \cite{Miura:2011mc} and with our model formula. 

 The proposed mechanism suggests  an approximate ``$N_f N_M$ scaling'', e.g. that the effect of adding $N_c$ fundamental fermions is about as large as adding one adjoint. Essentially for a given gauge group, the critical couplings shall fall on one single curve when plotted against the combination $ N_f N_M$ with various fermion representations and flavor numbers.  Such scaling can be readily tested.

5. To further ``probe'' the fermions' effect on confinement transition, one may turn on a quark chemical potential $\mu_q$ and see how the critical coupling changes accordingly. With the presence of a small $\mu_q$, there will be difference if a zero mode is occupied by a quark or an anti-quark. This effect can be incorporated into the present model by replacing $(1+f)^2$ in Eq.(\ref{Eq_beta}) by $(1+f e^{z})(1+f e^{-z})$ with $z\equiv \mu_q/T$. We then obtained:
\begin{eqnarray}
\frac{\beta_c(N_f,z)}{\beta_c(N_f,z=0)} =   1 - \frac{4 N_f N_M}{15} \frac{f}{(1+f)^2}\, z^2 + \hat{O}(z^4)  \quad
\end{eqnarray}
for small chemical potential $\mu_q<<T$. Present lattice simulations, though not capable of handing finite $\mu$ directly due to the sign problem, are actually able to extract such dependence (or Taylor coefficients in $\mu_q/T$ expansion) \cite{Kaczmarek:2011zz} and test the above prediction. Similar estimate can be made for  the dependence on the isospin chemical potential as well as the axial chemical potential.

 The existence of these monopole-quark states also implies a contribution from them to the thermal fluctuations of the conserved charges (e.g. baryon number) they carry, thus contributing to the quark number susceptibilities \cite{Borsanyi:2011sw,Liao:2005pa}. A monopole may have one, two, or more zero modes occupied, but since the extracted $f\approx 0.154$ is small the main contribution would be from the monopoles with one single zero-mode quark. The net baryonic density from these monopole-quark states (with the presence of a small quark chemical potential $z=\mu_q/T$) can be estimated as $n_{m-q}/T^3 \approx (n/T^3)\, (e^z-e^{-z}) f /[(1+fe^z)(1+fe^{-z})]^{N_f N_M}$. This yields a contribution to the quark number susceptibilities $\chi_i=\partial^{i-1}(n_{m-q}/T^3)/\partial (\mu_q/T)^{i-1}$ as
$\chi^{m-q}_2 \approx \frac{n}{T^3} \frac{2 f}{(1+f)^{N_f N_M}} \sim 0.4 - 0.8$
where we used lattice results for a total monopole density near $T_c$ to be $n/T^3\sim 2 - 4$, and $N_f N_M =3$. The so-obtained number makes a significant fraction of the lattice results around $T_c$ \cite{Borsanyi:2011sw}. Higher-order susceptibilities can be estimated similarly and the multi-quark-monopole states with higher charges may be important there. 
Similar effects can be estimated along this line also for the isospin and electric charge fluctuations.

6. It is well known that the $N_c=2$ theory is a very special case, with extra symmetry between quarks-antiquarks and mesons-diquarks. It also allows the finite density lattice simulations without
the ``sign problem''.
Lattice study of this theory was recently extended to the low-$T$ finite-$\mu$ region with $N_f=2,4$ quarks by Hands, et al \cite{Hands:2011ye}. 
The quark density (per flavor) shown in their Fig.3 displays a number of features:
(i) a structure at $\mu\approx m_\pi/2$ 
as predicted by the rotation from the $\bar{\psi}\psi$ to diquark condensate \cite{Shuryak:1998my,Kogut:2000ek};  (ii) the usual quark Fermi sphere at higher $\mu$; and (iii) an unexpected growth of quark density at still higher
$\mu$ to about twice the value as expected from the  Fermi sphere. The deconfinement as per the Polyakov loop appears concurrent with (iii).  

We now propose that this extra quark density in (iii) is due to the  condensate of the  {\em monopole-quark} states. The high quark chemical potential strongly favors states with quark numbers and  efficiently converts the pure monopoles into monopole-quark states, thus explaining the deconfinement at about the same density. At such low $T$ the dominant monopole-single-quark objects, being bosonic, would appear mostly as a condensate, like the diquarks. Assuming standard effective potential with  a repulsive binary interaction $V_{int}=\lambda n^2/4$, one gets the condensate density growing linearly with $\mu$, 
$n_{BEC}\sim   (\mu - m ) / \lambda   $
at $\mu>m$, which is consistent with observations of \cite{Hands:2011ye}.
Furthermore,  a similar density per flavor for both $N_f=2$ and  $N_f=4$ is
consistent with our view that these objects are dominantly states with only a $single$ quark per monopole. States with multiple quark may become relevant at even higher density though.  This proposal should and can be checked in many ways.  Direct monopole-flavor correlations can be seen in the configurations of these 
simulations.  One may also find the evaporation  
of this BEC as $T$ is above certain critical value, into the Bose gas of 
such monopole-quark states.

7. While we believe the main idea is  robust, the model here is admittedly crude and  intended to lead toward further studies and direct tests by dedicated lattice data. In particular, 
 the correlation between the monopole line with flavor-carrying
 fermionic operators can be used to measure the probability for quarks ``riding" on  monopoles.
 The discussion here is limited to the usual confinement, due to BEC of ``empty" monopoles. 
More exotic objects, bosonic monopoles $with$ quarks or even two monopoles bound by quarks,
can in principle undergo BEC as well: see recent discussion in \cite{Shuryak:2012ua}.

{\bf Note added}: After submitting the paper we became aware of  \cite{Agasian:2001an} discussing a similar problem in 2+1 dimensions. While the phase transition in this case is not of Bose condensation type and monopoles are substituted by vortices, the role of fermionic zero modes is similar and it also leads to reduced transition temperature.

 {\bf Acknowledgments.} 
The authors thank M.D'Elia and Ya. Shnir for discussions.  JL acknowledges support by RIKEN BNL Research Center. ES is supported in parts by the US-DOE grant
DE-FG-88ER40388.

\end{document}